\documentclass[preprint,12pt,number]{elsarticle}

\usepackage{url}
\usepackage{listings}
\usepackage{comment}
\usepackage{subcaption} 
\usepackage{amssymb}
\usepackage{amsmath}
\usepackage{booktabs}
\usepackage{xcolor}

\usepackage{pgfplots}
\usepgfplotslibrary{statistics}
\pgfplotsset{compat=1.18}

\usepackage{tikz}
\usetikzlibrary{arrows.meta, positioning, shapes.geometric, calc}

\journal{arXiv preprint}

\begin{document}

\begin{frontmatter}

\title{Assessing Problem-Solving in HR Contexts: A Comparison Between Game-Based and Self-Report Measures}

\author[unicam]{Fabrizio Fornari\corref{cor1}}
\ead{fabrizio.fornari@unicam.it}

\author[sapienza]{Eleonora Cova}
\author[unicam]{Niccol\`o Vito Vacca}
\author[sipi]{Francesco Bocci}
\author[bicocca]{Marcello Sarini}
\author[emaze]{Luigi Caputo}

\cortext[cor1]{Corresponding author}

\affiliation[unicam]{
  organization={University of Camerino},
  addressline={Computer Science Division, School of Science and Technology},
  city={Camerino},
  country={Italy}
}

\affiliation[sapienza]{
  organization={Sapienza University of Rome},
  addressline={Department of Psychology of Developmental and Socialization Processes},
  city={Rome},
  country={Italy}
}

\affiliation[sipi]{
  organization={Societ\`a Italiana di Psicoterapia Individuale -- Adleriana},
  city={Reggio Emilia},
  country={Italy}
}

\affiliation[bicocca]{
  organization={University of Milano-Bicocca},
  addressline={Department of Psychology},
  city={Milan},
  country={Italy}
}

\affiliation[emaze]{
  organization={Emaze Gaming Societ\`a Benefit SRL},
  city={Naples},
  country={Italy}
}

\begin{abstract}
Game-based assessments (GBAs) are increasingly adopted in recruitment contexts as tools to assess transversal skills through observable behavior. However, empirical evidence directly comparing game-based behavioral indicators with traditional self-report measures remains limited. This study adopts a method-comparison approach to explore the convergence between self-perceived and behaviorally enacted problem-solving competence, comparing a game-based assessment with the Problem Solving Inventory (PSI-B). 
Seventy-two participants completed both the PSI-B and a five-minute game-based problem-solving task, which classified performance into four behavioral proficiency levels. Results revealed no significant convergence between self-reported and behavior-based problem-solving scores, indicating a lack of convergence between the two measurement modalities. 
Rather than indicating a lack of validity of the game-based assessment, these findings support the view that self-report and behavioral measures provide complementary information about problem-solving competence. The study highlights the risks of relying on a single assessment modality in personnel selection and underscores the value of integrating game-based tools within multi-method assessment frameworks.
\end{abstract}


\end{frontmatter}

\section{Introduction}
\label{chap:Introduction}

Organizations today operate in an increasingly competitive labor market, often described as a global “war for talent,” in which identifying and attracting the right candidates has become a strategic priority \cite{woods2020}. This challenge is amplified by the growing demand for employees who possess strong transversal soft skills (e.g., problem-solving, teamwork, communication, adaptability), which have become critical predictors of employability and long-term success \citep{deming2017socialskills,de2020,bhati2022,schmitt2014}. 
Traditional recruitment and selection methods have long played a central role in assessing candidates’ competencies and remain widely adopted in organizational practice. However, several authors (e.g., Carless, 2009; Schmitt, 2014) have noted that these approaches may face challenges in keeping pace with evolving organizational needs and candidate expectations. Standard tools such as structured interviews, cognitive tests, and self-report questionnaires, while well-established and psychometrically grounded, have been criticized for being perceived as less engaging in contemporary recruitment contexts \cite{shankar2024}. Moreover, prior research has highlighted a number of potential limitations, including susceptibility to response distortion, social desirability bias, test anxiety, and constraints on ecological validity \cite{carless2009}. In particular, traditional assessment formats may struggle to fully capture the dynamic and situational nature of soft skills, which are often expressed through behavior rather than static trait indicators \cite{schmitt2014}. 

Among transversal soft skills, problem-solving represents a particularly relevant and methodologically challenging construct to assess, due to its dynamic, process-oriented, and context-dependent nature. 
Problem-solving is a dynamic, cognitively demanding process involving the ability to identify issues, analyze information, generate alternatives, and implement effective solutions—skills that unfold over time and within context-dependent situations \citep{Pouezevara2019,keen2011}. In workplaces, systematic problem-solving approaches tend to lead to better outcomes than intuitive methods \citep{tyre1993,karla2022}.
Conventional assessment formats, including structured interviews, stress interviews, and psychometric questionnaires, attempt to approximate these competencies but often fall short. Stress interviews, for example, seek to simulate pressure to observe candidates’ coping strategies, yet such methods risk eliciting negative emotional reactions and damaging perceptions of organizational fairness and attractiveness \cite{carless_imber2007}. Self-report instruments, such as the Problem Solving Inventory (PSI) \cite{heppner1982psi}, provide standardized measurement but rely heavily on candidates’ introspective accuracy and willingness to disclose, making them vulnerable to social desirability bias, response distortion, and test anxiety. Moreover, these tools conceptualize problem-solving primarily as a stable disposition, capturing how individuals believe they approach challenges rather than how they actually behave when confronted with complex or unexpected situations. This misalignment can influence the degree to which assessment outcomes mirror actual work-related problem-solving processes, increasing the risk of inaccurate hiring decisions, poor person–job fit, and long-term performance mismatches.

As organizations confront the limitations of traditional assessment tools, technology-enhanced recruitment methods have emerged as potential alternatives capable of addressing both organizational and candidate expectations. Recent integrative reviews have highlighted how recruitment research is increasingly shaped by emerging digital tools and have called for further empirical investigation into technology-based assessment approaches \cite{sharma2026}. Among these innovations, Game-Based Assessments (GBAs) have gained considerable attention for their ability to create more engaging, realistic, and behaviorally rich evaluation environments \cite{marczewski2015}. A key advantage of GBAs lies in their capacity to enhance the overall recruitment experience. By incorporating interactive and immersive elements, organizations can signal technological sophistication and modernity, thereby strengthening employer attractiveness and differentiating themselves in the competitive talent landscape \cite{gkorezis2024}. 

Beyond their experiential appeal, GBAs offer substantive methodological advantages. Unlike traditional self-report instruments, game-based assessments capture direct behavioral data as individuals interact with dynamic tasks. This facilitates the assessment of transversal competencies through the strategies, choices, and actions displayed during gameplay, rather than through self-descriptions or static test items \cite{armstrong2016}. Such behavioral evidence is particularly valuable for evaluating soft skills, which are inherently context-dependent and often poorly reflected in conventional assessment formats.

Therefore, gamified environments have been shown to mitigate several sources of bias commonly associated with traditional testing. By embedding challenges within engaging, task-focused activities, GBAs can reduce test anxiety, stereotype threat, and social desirability bias, fostering more authentic behavioral expressions \citep{mcpherson2008,albuquerque2017}. When candidates focus on task completion rather than impression management, the likelihood of faking decreases, contributing to fairer and more reliable assessments. 

This raises a broader measurement question as to whether different assessment modalities tap into the same underlying construct or capture distinct, yet related, facets of problem-solving competence. In the context of personnel selection, addressing this question is particularly relevant, as reliance on a single assessment modality may lead to incomplete or biased representations of candidates’ capabilities

The present study examines a Game-Based Assessment (GBA), hereafter referred to as Behaveme-PS
, which operationalizes problem-solving competence through observable in-game behavior. Rather than treating traditional psychometric instruments as criterion measures, the study adopts a method-comparison approach to investigate the relationship between behavioral indicators derived from gameplay and scores obtained through a widely used self-report instrument, the Problem Solving Inventory (PSI). From a theoretical perspective, this comparison allows us to empirically explore the degree of convergence or divergence between self-perceived and behavior-based operationalizations of problem-solving, addressing an open question in the personnel selection and GBA literature. By documenting how these two measurement modalities relate—or fail to relate—this study contributes to a more nuanced understanding of the role of GBAs as complementary tools, rather than interchangeable alternatives, for the evaluation of transversal skills in organizational settings.

The paper first reviews the relevant literature, then describes the study design and analytical approach, and presents the main findings. The final sections discuss the results, their implications for research and HR practice, and outline directions for future work.

\section{Literature Review}
\label{sec:related}

\subsection{Problem-Solving Soft Skill}

Soft skills encompass a broad range of interpersonal, cognitive, and self-regulatory abilities that enable individuals to interact effectively and adapt to dynamic environments. Unlike hard skills, which involve technical expertise related to specific tasks (e.g., programming, data analysis), soft skills capture how individuals navigate social, emotional, and problem-oriented demands. Competencies such as problem-solving, emotional intelligence, communication, teamwork, persistence, and leadership are widely recognized as essential for effective functioning in contemporary workplaces, particularly in collaborative contexts where communication, coordination, and adaptability directly influence team performance \citep{karimi2021,de2020,bhati2022}.

Within this broader domain, problem-solving is frequently identified as a core transversal skill relevant across professional roles and industries. Problem-solving is generally defined as the ability to acquire or use prior knowledge to address novel challenges \cite{Pouezevara2019}. It develops from early childhood and continues to shape performance in adulthood, underpinning individuals’ capacity to analyze situations, generate solutions, and make informed decisions \cite{keen2011}. Effective problem-solving typically involves breaking down complex issues into manageable components, evaluating alternative courses of action, and implementing the most appropriate solution. Research shows that systematic approaches to problem-solving tend to yield more effective outcomes than purely intuitive strategies \cite{tyre1993}.

Its classification as a ``transversal'' skill underscores its applicability across diverse contexts and tasks \cite{karla2022} with skilled problem-solvers consistently demonstrating superior abilities in identifying issues, assessing their severity, and selecting appropriate strategies.

Despite their acknowledged importance, soft skills, in particular problem-solving, remain inherently difficult to assess, as they are expressed primarily through behavior and situational responses rather than through declarative or technical knowledge. This characteristic poses significant challenges for traditional assessment approaches and has important implications for recruitment and selection practices.

\subsection{Traditional Psychometric Approaches}

Psychometric tests constitute one of the most established and widely used approaches for assessing soft skills in scientific research and Human Resources (HR) practice. These instruments can measure cognitive abilities, personality traits, attitudes, and behavioral tendencies, and have been applied across a range of professional and academic domains \cite{kumar2023}. In the context of soft-skill evaluation, standardized psychometric tools have been developed to assess constructs such as problem-solving, emotional intelligence, and interpersonal skills in a reliable and validated manner \citep{jardim2022,tountopoulou2021}. Their extensive adoption in HR is largely driven by their standardized administration and scoring procedures, which ensure consistency, objectivity, and the generation of quantifiable data to support decision-making across recruitment, promotion, and talent development processes \citep{carless2009,kline2013}.

A key advantage of psychometric assessments is their standardized administration and scoring procedures, which ensure that all participants are evaluated using the same criteria. This standardization enhances objectivity and consistency across candidates, reducing some forms of evaluator bias and enabling data-driven decisions that can complement interviews or résumé screening \citep{kline2013,carless2009}. Because many soft skills are difficult to observe directly in applied settings \cite{deepa2013}, structured psychometric assessments offer HR professionals a means to evaluate these attributes in a controlled and comparable format. Typically administered online or in paper-based formats and interpreted by trained professionals, these instruments are widely used in recruitment, performance appraisal, and leadership development programs \cite{cripps2017}. However, despite the advantages, psychometric tests, particularly self-report measures, present several well-documented limitations. 

Problem-solving, in particular, has been extensively examined within psychometric frameworks due to its relevance across a wide variety of professional tasks. Among the tools available, the \emph{Problem Solving Inventory} (PSI) is one of the most established instruments for assessing individuals’ perceived problem-solving abilities \cite{heppner1982psi}. The PSI captures multiple dimensions of problem-solving, including confidence in handling challenges, preferred coping strategies, and tendencies toward avoidant or active approaches. As a self-report measure, it provides insight into how individuals believe they respond to stress, uncertainty, and complex decision-making demands, rather than how they behave in real-time workplace situations.

Despite their widespread use, self-report psychometric instruments such as the PSI are vulnerable to several sources of bias. Social desirability, response distortion, and faking are particularly salient in high-stakes contexts such as personnel selection, where candidates can easily infer desirable traits and strategically adjust their responses. Cultural bias remains a concern, as some instruments include items that may favor individuals from specific cultural or linguistic backgrounds, potentially undermining fairness in diverse organizational settings \cite{vandervijver2004}. Moreover, formal testing situations can induce test anxiety, which may impair performance on cognitively demanding tasks such as problem-solving, further reducing ecological validity \cite{Zeidner1998TestAnxiety}.
As a result, while psychometric tools effectively capture broad dispositional tendencies, they often struggle to reflect the dynamic, context-dependent behaviors that characterize real-world cognitive and social performance.
Taken together, these limitations have prompted growing interest in assessment approaches that move beyond self-report measures and static testing formats by capturing behavior in dynamic, context-rich environments.

\subsection{Game-Based Assessments} 
As part of broader efforts to expand and refine assessment approaches, research on digital assessment has increasingly distinguished between gamification, serious games, and game-based assessments (GBAs).
Gamification involves adding game elements (e.g., points, badges, or leaderboards) to traditional tests to enhance engagement, without altering their underlying reliance on self-reported responses \cite{Deterding2011Gamification}. Serious games, by contrast, are full game environments designed for purposes beyond entertainment, offering immersive and interactive scenarios that can support learning and behavioral measurement \citep{MichaelChen2006SeriousGames,Connolly2012SeriousGamesReview}. 
GBAs represent a specialized subset of serious games in which assessment is embedded directly within gameplay, enabling real-time observation of behavior as individuals interact with dynamic tasks. Unlike gamification or general serious games, GBAs are explicitly designed for measurement purposes, capturing behavioral indicators of underlying psychological constructs through immersive challenges and automated data collection \citep{ShuteKim2014StealthAssessment,Shute2009EmbeddedAssessment}. 

A defining characteristic of GBAs is that measurement is embedded within naturalistic interaction with dynamic tasks, allowing the observation of real-time behavior as individuals engage with complex problem-solving situations. This design supports the collection of rich behavioral data and enhances ecological validity by approximating the contextual demands under which competencies such as problem-solving are enacted \citep{seo2018,altomari2023,khaderi2025,xuli2022}. By capturing both explicit, controlled decision-making and more implicit, automatic processes, GBAs can provide a nuanced representation of underlying competencies that is difficult to obtain through self-report measures alone.

From a measurement perspective, prior research suggests that GBAs can achieve acceptable levels of fairness, reliability, and predictive accuracy \citep{GEORGIOU2020106356,altomari2023}. Stealth assessment mechanisms embedded in gameplay have been associated with reduced test anxiety while maintaining measurement quality \cite{altomari2022}, and reliance on observed behavior rather than explicit self-descriptions may reduce vulnerability to intentional faking and certain rater-based biases \citep{xuli2022,armstrong2016,ramos2022}. In addition, because assessment is embedded within game interaction, candidates may perceive GBAs as less overtly evaluative, which has been associated with higher engagement and the expression of more naturalistic behavioral responses \citep{altomari2022,altomari2023,negron2022,xuli2022}. At a process level, several authors also note that game-based e-recruitment solutions may support efficiency and scalability in early stages of HR assessment, facilitating the management of large applicant pools without replacing more in-depth evaluation methods \citep{Georgiou2019,altomari2022}.

Because of these characteristics, GBAs have been proposed as a behavior-based approach for assessing complex soft skills in HR contexts. Illustrative examples include Firefly Freedom and Use Your Brainz, which derive problem-solving indicators from gameplay behavior \citep{consultancyuk2015,xuli2022,shute2016}. 

Despite the growing body of work on game-based assessments, the literature discusses a range of open issues related to their development and use, including questions of fairness, inclusivity, design complexity, and candidate perceptions \citep{khaderi2025,xuli2022, armstrong2016,Georgiou2019,bylieva2018}. Within this broader research landscape, one issue that is particularly relevant for the interpretation of game-based assessment outcomes is the scarcity of empirical studies directly comparing gameplay-derived behavioral indicators with established psychometric measures \citep{marengo2023ile,shute2016}. Addressing this gap is especially important for complex competencies such as problem-solving, where self-reported perceptions and enacted behavior may capture different facets of the same construct. The present study directly contributes to this line of research by examining the relationship between a game-based assessment of problem-solving and a widely used self-report instrument, the Problem Solving Inventory (PSI).

\section{Methodology}
\label{sec:methodology}

\subsection{Procedure}
This cross-sectional method-comparison study examines the convergence between self-perceived and behaviorally enacted problem-solving competence by comparing a game-based assessment with a traditional self-report measure.
Specifically, we compared the results obtained from Behaveme-PS with those of the Problem Solving Inventory (PSI-B) \cite{heppner1982psi}.
This setup allows for a direct method comparison, in which PSI-B scores are examined alongside behavioral proficiency levels derived from gameplay.  
To standardize the procedure and minimize cognitive load on participants, the entire workflow was administered through a single Microsoft Form, which served as the unified interface for instructions, test administration, and data entry. The full process lasted approximately 20 minutes and consisted of two sequential phases. 
The first phase of the study consisted of administering the PSI-B, which participants completed before engaging with the Behaveme-PS module. In accordance with professional standards for psychological assessment, the PSI-B was administered under the supervision of a licensed psychologist. 
Participants completed the PSI-B digitally through the Microsoft Form used for the study, and their responses were automatically recorded. This created a structured psychometric dataset that served as the reference self-report dataset for comparison with the behavioral indicators obtained from the Behaveme-PS assessment. PSI-B scores were categorized into four levels (Base, Medium, High, Very High) using a quartile-based classification to ensure distribution-sensitive and non-arbitrary group assignment.
Upon completing the PSI-B, participants were redirected to the Behaveme platform to complete a single 5-minute session of Behaveme-PS. During gameplay, the system captured several behavioral indicators, including the total number of valid combinations completed, the overall number of moves, and the average time required per combination. At the end of the session, Behaveme-PS provided each participant with a performance classification along a four-level scale (Base, Medium, Advanced, Very Advanced). 
Participants then manually entered the summary performance classification provided by Behaveme-PS into the Microsoft Form, allowing the dataset to pair each individual’s psychometric profile (PSI-B) with their behavioral performance. 
Data management, and statistical analyses were conducted independently by the academic research team. Emaze Gaming Società Benefit SRL, the company that developed the Behaveme-PS platform, did not participate in data processing or statistical analyses, and had no role in the interpretation or reporting of the findings.

\subsection{Participants}
A total of seventy-eight participants initially took part in the study. 
All participants were undergraduate or graduate students enrolled in the Computer Science curricula at the University of Camerino (UNICAM), were aged 18 years or older, and volunteered to participate. Participants were recruited during scheduled class sessions. Participation was entirely optional, and no academic credit or material incentives were offered.

The privacy and rights of all human subjects were fully respected. Prior to participation, all individuals provided informed consent. The study received formal approval from an institutional ethical committee and was conducted in full compliance with ethical standards for research involving human participants. 

For the purposes of analysis, only results from participants who provided complete responses on the self-report questionnaire and reported the result of the Behaveme-PS were retained, resulting in a final sample of seventy-two participants with complete data. 

\subsection{Measures}
\textbf{Problem Solving Inventory}. 
To assess self-perceived problem-solving competence, several validated psychometric instruments have been developed. Among these, the Problem Solving Inventory (PSI; Cronbach’s $\alpha = .92$) is one of the most widely used and empirically supported measures. Originally developed by \cite{heppner1982psi}, the PSI assesses individuals’ perceptions of their problem-solving abilities rather than their actual performance.

In the present study, we employed the PSI-B, a revised version of the original instrument adapted for contemporary assessment contexts, which preserves the core theoretical structure of the PSI while offering updated item wording and response formatting. The PSI-B captures individuals’ general beliefs, confidence, and perceived control regarding how they typically approach problem-solving situations (e.g., “Usually, I am able to think of intelligent and effective solutions to my problems”), rather than task-specific performance outcomes (e.g., “After taking certain action to solve a problem, I compare the results obtained with those I had predicted”). Participants rated their level of agreement with each item on a six-point Likert scale ranging from 1 (strongly agree) to 6 (strongly disagree).

\textbf{Behaveme - Problem Solving.} Behaveme is a digital platform, developed by the Emaze Gaming company,\footnote{\url{https://www.emazegaming.it/}} to deliver game-based assessments (GBAs) for use in personnel selection and talent development. The platform hosts several assessment games targeting different soft skills; in this study, we focus specifically on the Problem-Solving module. 
The Problem-Solving game available from the Behaveme platform, that we refer with the term Behaveme-PS, simulates real-world cognitive demands through an interactive task in which candidates reorganize elements under time constraints. The design of the game draws on principles from two established cognitive assessment tools: the Raven’s Progressive Matrices \cite{raven2000}, which assesses fluid intelligence and pattern recognition and the Tower of Hanoi Test \cite{humes1997towers}, which measures planning and sequential problem-solving ability. These psychometric foundations are embedded in gameplay. 

\begin{figure*}[t]
\centering
\subfloat[Main game interface.\label{fig:game_start}]{
  \includegraphics[width=0.475\linewidth]{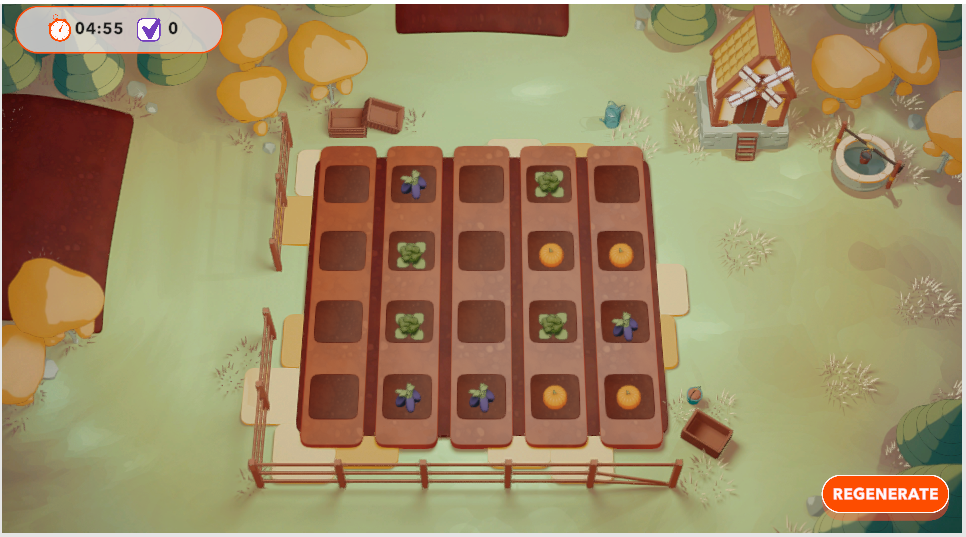}
}
\hfill
\subfloat[Game result screen.\label{fig:game_end}]{
  \includegraphics[width=0.475\linewidth]{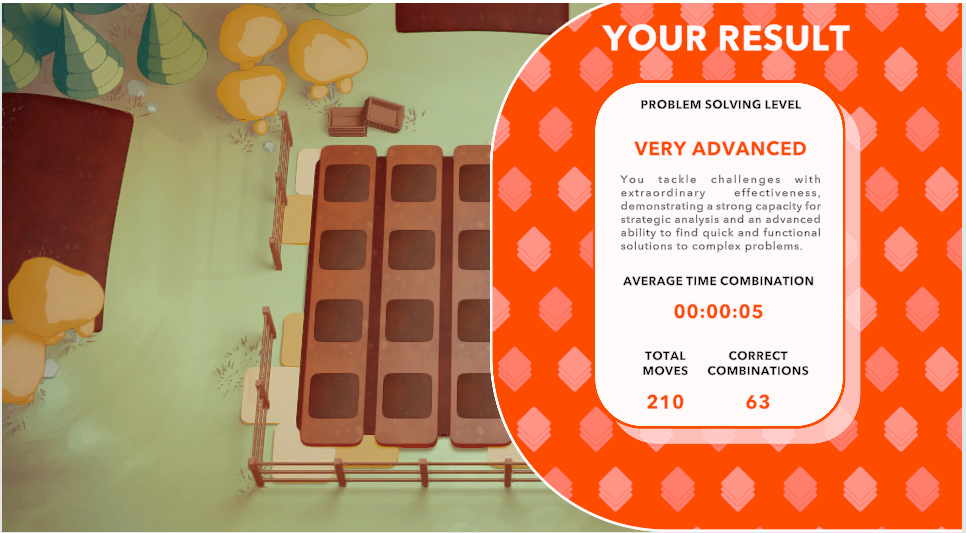}
}
\caption{Screenshots of the game-based problem-solving assessment used in the study.}
\label{fig:game}
\end{figure*}

Screenshots of the Behaveme-PS assessment used in the study are reported in Figure \ref{fig:game}. 
The gameplay experience is structured to elicit behaviors that map onto three fundamental dimensions of problem-solving. The first is strategic planning, which emerges as players coordinate limited space and resources, deciding how to reorganize elements efficiently while minimizing unnecessary actions. The second dimension, sequential thinking, is activated as participants must anticipate how early moves will influence later steps, often requiring them to visualize several transitions ahead before committing to an action. Finally, adaptability plays a central role: because the game offers minimal instruction and imposes time pressure, players must quickly infer the rules, adjust their strategies, and refine their approach as new situations unfold. 
These dimensions are not measured through self-report but through observable gameplay behavior. To capture them, the system records indicators such as the total number of combinations successfully completed, the number of moves taken, and the average time required to complete each combination. Together, these metrics provide a behavioral portrait of how each player approaches a novel, time-limited cognitive task, offering insight into their efficiency, cognitive flexibility, and decision-making style.

At the end of the 5-minute session, the platform summarizes the player’s performance by assigning them to one of four levels of problem-solving competence. These levels are designed to translate raw gameplay data into an interpretable profile. Players in the Base category tend to rely on immediate, trial-and-error solutions and may struggle to develop broader strategies. Those classified as Medium typically show a more balanced approach, demonstrating the ability to evaluate alternatives and select functional solutions, though sometimes without full optimization. Participants reaching the Advanced level generally exhibit clear planning, efficient execution, and strong analytical skills, often engaging with the challenges in a structured and purposeful manner. Finally, the Very Advanced category reflects highly optimized performance, where players demonstrate rapid comprehension of the task, excellent strategic reasoning, and the ability to generate effective solutions even under time pressure.

\subsection{Analysis}
Statistical analyses were conducted using Jamovi\footnote{\url{jamovi.org}} (version 2.6.26.0). Descriptive statistics were first computed for all study variables. Scores obtained on the PSI-B were initially treated as a continuous measure of self-perceived problem-solving ability \cite{kourmousi2016validity}. To allow a direct comparison with the categorical competence levels produced by the Behaveme-PS, PSI-B scores were subsequently divided into quartiles, yielding four ordinal categories (Base, Medium, High, Very High).
The quartile-based classification was adopted for both methodological and theoretical reasons. From a methodological perspective, it allowed alignment between a continuous self-report measure and a discrete, behavior-based assessment. 
From a theoretical standpoint, this approach is consistent with a method-comparison perspective, as it enables the comparison of perceived problem-solving competence with observed behavioral performance while reducing within-group variability and enhancing interpretability.
Moreover, given the absence of standardized cut-off scores for the PSI, a data-driven quartile classification represents a transparent and replicable strategy.
To examine the association between self-perceived and behavior-based problem-solving, a Spearman’s rank-order correlation was computed. This non-parametric approach was selected due to the ordinal nature of the categorized variables and the potential violation of normality assumptions.
To further explore how behavioral performance was distributed across levels of self-reported problem-solving competence, a contingency table was examined.

\section{Findings}
\label{sec:findings}
To examine the relationship between self-reported and behaviorally assessed problem-solving competence, we analyzed the association between PSI-B quartiles and Behaveme problem-solving levels using ordinal correlation and descriptive analyses.
A Spearman correlation analysis revealed no significant association between Behaveme-PS levels and PSI-B scores ($\rho = -.09, p = .45$), indicating a lack of monotonic relationship between behavioral and self-reported problem-solving competence (see Table \ref{tab1}).

\begin{table}[ht]
\centering
\caption{Correlations between PSI-B and Behaveme.}
\begin{tabular}{llcc}
\hline
 &  & \textbf{PSI-B} & \textbf{Behaveme} \\
\hline
\textbf{PSI-B} 
& Spearman’s $\rho$ & -- & -- \\
& df             & -- & -- \\
& p-value        & -- & -- \\
\hline
\textbf{Behaveme} 
& Spearman’s $\rho$ & -0.090 & -- \\
& df             & 70     & -- \\
& p-value        & 0.452  & -- \\
\hline
\end{tabular}
\label{tab1}
\vspace{0.5em}

\begin{flushleft}
\footnotesize \textit{Note.} * $p < .05$, ** $p < .01$, *** $p < .001$
\end{flushleft}
\end{table}

A contingency table was used to examine how behavioral performance varied across levels of self-reported problem-solving competence (see Table 2). The distribution revealed substantial variability within PSI-B quartiles and no clear alignment between self-reported and behaviorally assessed problem-solving levels.

\begin{table}[ht]
\centering
\caption{Distribution of Behaveme levels across PSI-B quartiles}
\begin{tabular}{lcccc}
\toprule
& \multicolumn{4}{c}{\textbf{Behaveme Level}} \\
\cmidrule(lr){2-5}
\textbf{PSI-B Quartile} 
& \textbf{Base} 
& \textbf{Medium} 
& \textbf{Advanced} 
& \textbf{Very advanced} \\
\midrule
\textbf{Base}      & 1 & 7 & 7 & 3 \\
\textbf{Medium}    & 3 & 5 & 6 & 4 \\
\textbf{High}      & 5 & 5 & 7 & 2 \\
\textbf{Very High} & 5 & 3 & 5 & 4 \\
\bottomrule
\end{tabular}
\label{tab2}

\vspace{0.5em}
\begin{flushleft}
\footnotesize \textit{Note.} Values represent frequency counts.
\end{flushleft}
\end{table}

\begin{figure}[htbp]
\centering
\begin{tikzpicture}
\begin{axis}[
    width=\linewidth,
    height=0.60\linewidth,
    ymin=0.8, ymax=4.2,
    ytick={1,2,3,4},
    xtick={1,2,3,4},
    xlabel={Behaveme},
    ylabel={PSI-B},
    boxplot/draw direction=y,
    boxplot/box extend=0.3,
    boxplot/every box/.style={fill=blue!25, draw=black},
    boxplot/every median/.style={very thick, black},
    boxplot/every whisker/.style={black},
    boxplot/every cap/.style={black},
]
\addplot+[
  boxplot prepared={
    lower whisker=1,
    lower quartile=2.25,
    median=3,
    upper quartile=4,
    upper whisker=4
  }
] coordinates {};

\addplot+[
  boxplot prepared={
    lower whisker=1,
    lower quartile=1,
    median=2,
    upper quartile=3,
    upper whisker=4
  }
] coordinates {};

\addplot+[
  boxplot prepared={
    lower whisker=1,
    lower quartile=1,
    median=2,
    upper quartile=3,
    upper whisker=4
  }
] coordinates {};

\addplot+[
  boxplot prepared={
    lower whisker=1,
    lower quartile=2,
    median=2,
    upper quartile=4,
    upper whisker=4
  }
] coordinates {};
\end{axis}
\end{tikzpicture}
\caption{Distribution of PSI-B scores across the four behavioral groups.}
\label{fig:psib-boxplot}
\end{figure}
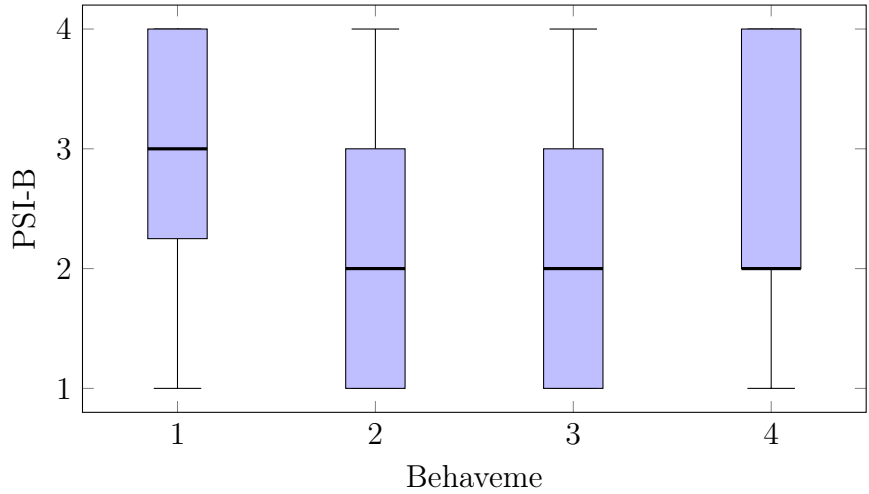

Consistent with Table \ref{tab2}, the box plot (Figure \ref{fig:psib-boxplot}) illustrates considerable within-group variability in PSI-B scores across behavioral performance levels, with largely overlapping distributions and no clear monotonic relationship between self-reported problem-solving competence and behaviorally assessed performance.

To summarize, the data revealed a complex pattern. Not all participants classified as Very High based on the psychometric assessment achieved correspondingly high levels in the game-based evaluation. Conversely, several participants classified in the Medium or High PSI-B quartiles demonstrated behavioral performance levels comparable to those observed in the Advanced group.
In particular, participants in the Very High PSI-B quartile were distributed across multiple Behaveme-PS levels, indicating marked within-group heterogeneity. While some individuals reached the Very Advanced level, the majority were classified at lower behavioral levels. Similarly, participants in the High and Medium PSI-B quartiles were frequently observed in higher Behaveme-PS categories, suggesting that strong behavioral problem-solving performance was not exclusive to those reporting the highest levels of self-perceived competence. Participants in the Base PSI-B quartile were not predominantly classified in lower Behaveme-PS categories; rather, with the exception of a single participant, most were distributed across Medium to Very Advanced behavioral performance levels, indicating substantial heterogeneity within this group.

\section{Discussion}
\label{sec:discussion}

The present study aimed to compare a traditional self-report measure (PSI-B) with a game-based assessment, examining the degree of convergence between self-perceived and behaviorally enacted problem-solving. The results revealed no significant association between PSI-B scores and behavioral performance levels derived from gameplay. 

The lack of convergence observed between the PSI-B and the game-based assessment is consistent with a growing body of evidence indicating that self-report and behavioral measures of the same psychological construct are often weakly associated \citep{dang2020,shute2015}. While the PSI-B captures individuals’ beliefs, confidence, and self-evaluations regarding how they typically approach problems, the game-based assessment operationalizes problem-solving as a dynamic behavioral process unfolding in real time under task constraints. These two measurement approaches rely on fundamentally different response processes: self-report measures require reflective judgments aggregated across diverse and largely unstructured real-life contexts, whereas behavioral measures elicit performance in highly structured, time-limited, and uncommon situations. Moreover, behavioral tasks tend to capture maximal performance under optimal effort conditions, while self-report instruments more closely reflect typical performance and subjective perceptions of competence. As a result, the two measures may index partially distinct facets of problem-solving, rather than interchangeable indicators of a single underlying construct \cite{borgonovi2023}. These findings contribute to ongoing debates regarding the construct validity and measurement equivalence of technology-mediated assessment tools in organizational psychology \cite{ramos2022}.

Further insight into this divergence is provided by the distribution of behavioral performance across levels of self-reported problem-solving competence. The contingency analysis revealed substantial within-group heterogeneity, with participants reporting high or very high problem-solving competence distributed across all levels of game-based performance. Conversely, several individuals reporting lower self-perceived competence demonstrated medium to advanced behavioral performance in the GBA. This pattern can be interpreted through the lens of the competence–performance discrepancy \cite{dang2020}, whereby self-report measures primarily capture perceived competence—individuals’ beliefs about their general ability to solve problems—whereas behavioral performance reflects the enactment of that competence in a specific task context. Importantly, behavioral performance is shaped not only by underlying ability, but also by situational factors such as motivation, effort, engagement, and task-specific demands \cite{kajonius2020}. As a result, high perceived competence does not necessarily translate into high behavioral performance in a constrained game-based environment, nor does lower self-perceived competence preclude effective problem-solving when situational conditions are favorable.

The observed divergence between self-reported and behavior-based measures has direct implications for how problem-solving competence should be assessed in personnel selection contexts. From an applied perspective, these findings suggest that, in high-stakes recruitment settings, exclusive reliance on self-report instruments may lead to the misclassification of candidates, resulting in false positives or false negatives with respect to actual problem-solving performance. GBAs can complement traditional tools by capturing how candidates behave when confronted with novel, time-pressured tasks, thereby providing additional information that may not emerge through self-report alone. Rather than replacing established psychometric instruments, GBAs may be most effective when integrated within multi-method assessment frameworks that combine perceived and behavioral indicators of competence.

The findings of the present study should be interpreted in light of several limitations. First, the sample was relatively homogeneous, consisting predominantly of young university students, which may limit the generalizability of the results. Although the sample consisted of university students, this population represents a relevant early-career cohort frequently involved in digital recruitment pipelines, making it a meaningful context for examining technology-mediated assessment tools. Future research should include more diverse and professionally heterogeneous samples to assess whether similar patterns emerge across different populations and career stages. Second, the study was conducted in an academic setting rather than in an applied organizational context. As problem-solving behavior is highly context-dependent, future studies should examine the predictive and ecological validity of GBAs in real-world recruitment and selection environments. 
Finally, individual difference variables such as self-efficacy, personality traits, motivation, prior gaming experience, and gender-related effect were not included in the present study. Incorporating these factors in future research may help clarify the mechanisms underlying the observed divergence between self-reported and behavior-based measures of problem-solving competence and support the development of more integrated, multi-method assessment approaches.

\section{Conclusion}
\label{sec:conclusion}

This study compared a traditional self-report measure (PSI-B) with a game-based assessment (Behaveme-PS) to examine the relationship between self-perceived and behaviorally enacted problem-solving skill. The findings revealed a lack of convergence between self-report and behavioral indicators, suggesting that these assessment modalities capture distinct facets of problem-solving and cannot be used interchangeably. This divergence underscores the importance of distinguishing between perceived competence and enacted performance when assessing complex, context-dependent skills. While self-report measures provide insight into individuals’ beliefs and self-evaluations, game-based assessments capture behavior as it unfolds under dynamic task constraints. Accordingly, integrating multiple assessment approaches may offer a more comprehensive and nuanced understanding of problem-solving competence. 

\subsection*{Author Contributions}

\textbf{Fabrizio Fornari}: Conceptualization, Methodology, Formal analysis, Supervision, Writing – Original Draft, Writing – Review \& Editing.  

\textbf{Eleonora Cova}: Interpretation of results, Writing – Original Draft, Writing – Review \& Editing.  

\textbf{Niccolò Vito Vacca}: Conceptualization, Methodology, Investigation, Data Curation.  

\textbf{Francesco Bocci}: Conceptualization, Investigation, Data Curation, Supervision, Interpretation of results.  

\textbf{Marcello Sarini}: Conceptualization, Interpretation of results.  

\textbf{Luigi Caputo}: Conceptualization (early-stage study design), Resources.

\subsection*{Conflict of Interest} 
Luigi Caputo is the founder of Emaze Gaming, the company that developed the game-based assessment platform used in this study. While he contributed to the conceptualization and methodological design of the study, he had no role in data collection, data analysis, interpretation of results, or reporting of findings. No financial compensation or funding was provided. The remaining authors declare no competing interests.

\subsection*{Ethics Statement}

This study was conducted in accordance with the ethical standards of the institutional research committee and the principles of the Declaration of Helsinki. Ethical approval was obtained from the University of Camerino Ethics Committee. All participants provided informed consent prior to participation. Participation was voluntary. All data were anonymized prior to analysis.

\subsection*{Funding}
This research did not receive any specific grant from funding agencies in the public, commercial, or not-for-profit sectors.

\subsection*{Data Availability}
The data that support the findings of this study will be made available upon reasonable request after publication.

\subsection*{Use of Artificial Intelligence}

The authors used an AI-based language model (ChatGPT, OpenAI) solely for language editing and formatting support. All scientific content, study design, data analyses, interpretations, and conclusions were developed independently by the authors and have been carefully reviewed and verified for accuracy.

\bibliographystyle{elsarticle-harv} 
\bibliography{biblio}

@article{sharma2026,
  title={Integrative Review of Recruitment Literature: Conceptual Evolution, Technological Developments, and Scope for Future Research},
  author={Sharma, Preeti and Padhi, Mousumi},
  journal={Human Behavior and Emerging Technologies},
  volume={2026},
  number={1},
  pages={3210624},
  year={2026},
  publisher={Wiley Online Library},
  doi= {10.1155/hbe2/3210624}
}

@article{albuquerque2017,
  author    = {Albuquerque, J. and Bittencourt, I. I. and Coelho, J. A. and Silva, A. P.},
  title     = {Does gender stereotype threat in gamified educational environments cause anxiety? An experimental study},
  journal   = {Computers \& Education},
  year      = {2017},
  volume    = {115},
  pages     = {161--170},
  doi       = {10.1016/j.compedu.2017.08.005}
}

@article{borgonovi2023,
  title={Gender differences in collaborative problem-solving skills in a cross-country perspective.},
  author={Borgonovi, Francesca and Han, Seong Won and Greiff, Samuel},
  journal={Journal of Educational Psychology},
  volume={115},
  number={5},
  pages={747--766},
  year={2023},
  publisher={American Psychological Association},
  doi = {10.1037/edu0000788}
}

@article{dang2020,
  author    = {Dang, J. and King, K. M. and Inzlicht, M.},
  title     = {Why are self-report and behavioral measures weakly correlated?},
  journal   = {Trends in Cognitive Sciences},
  year      = {2020},
  volume    = {24},
  number    = {4},
  pages     = {267--269},
  doi       = {10.1016/j.tics.2020.01.007}
}

@incollection{gkorezis2024,
  author    = {Gkorezis, P. and Georgiou, K. and Nikolaou, I. and Kyriazati, A.},
  title     = {Gamified or traditional situational judgement test? A moderated mediation model of recommendation intentions via organizational attractiveness},
  booktitle = {Recent Developments in Recruitment and Selection},
  year      = {2024},
  pages     = {82--92},
  publisher = {Routledge}
}

@article{kajonius2020,
  author    = {Kajonius, P. J. and Bj{\"o}rkman, T.},
  title     = {Individuals with dark traits have the ability but not the disposition to empathize},
  journal   = {Personality and Individual Differences},
  year      = {2020},
  volume    = {155},
  pages     = {109716},
  doi       = {10.1016/j.paid.2019.109716}
}

@book{marczewski2015,
  author    = {Marczewski, Andrzej},
  title     = {Even Ninja Monkeys Like to Play},
  year      = {2015},
  publisher = {Blurb Inc.},
  address   = {London}
}

@article{mcpherson2008,
  author    = {McPherson, J. and Burns, N. R.},
  title     = {Assessing the validity of computer-game-like tests of processing speed and working memory},
  journal   = {Behavior Research Methods},
  year      = {2008},
  volume    = {40},
  number    = {4},
  pages     = {969--981},
  doi       = {10.3758/BRM.40.4.969}
}

@inproceedings{schmitt2014,
  author    = {Schmitt, U.},
  title     = {The role of personal knowledge management systems in making citizens highly knowledgeable},
  booktitle = {INTED2014 Proceedings},
  year      = {2014},
  pages     = {2005--2014},
  publisher = {IATED}
}

@article{carless_imber2007,
  author  = {Carless, S. A. and Imber, A.},
  title   = {The influence of perceived interviewer and job and organizational characteristics on applicant attraction and job choice intentions: The role of applicant anxiety},
  journal = {International Journal of Selection and Assessment},
  year    = {2007},
  volume  = {15},
  number  = {4},
  pages   = {359--371},
  doi     = {10.1111/j.1468-2389.2007.00395.x}
}

@article{shankar2024,
  author    = {Shankar, C. S. U. and Ganesan, J.},
  title     = {Examining the Role of Gamification in Enhancing Candidate Engagement and Experience Through Gamified Selection Processes},
  journal   = {Pakistan Journal of Life \& Social Sciences},
  year      = {2024},
  volume    = {22},
  number    = {2},
  doi       = {10.57239/PJLSS-2024-22.2.001071}
}

@article{woods2020,
  author    = {Woods, S. A. and Ahmed, S. and Nikolaou, I. and Costa, A. C. and Anderson, N. R.},
  title     = {Personnel selection in the digital age: A review of validity and applicant reactions, and future research challenges},
  journal   = {European Journal of Work and Organizational Psychology},
  year      = {2020},
  volume    = {29},
  number    = {1},
  pages     = {64--77},
  doi       = {10.1080/1359432X.2019.1681401}
}

@incollection{Shute2009EmbeddedAssessment,
  author    = {Shute, Valerie J. and Ventura, Matthew and Bauer, Malte and Zapata-Rivera, Diego},
  title     = {Melding the Power of Serious Games and Embedded Assessment to Monitor and Foster Learning},
  booktitle = {Serious Games: Mechanisms and Effects},
  editor    = {Ritterfeld, Ute and Cody, Michael J. and Vorderer, Peter},
  year      = {2009},
  pages     = {295--321},
  publisher = {Routledge},
  address   = {New York},
}

@incollection{ShuteKim2014StealthAssessment,
  author    = {Shute, Valerie J. and Kim, Yanghee J.},
  title     = {Formative and Stealth Assessment},
  booktitle = {Handbook of Research on Educational Communications and Technology},
  editor    = {Spector, J. Michael and Merrill, M. David and Elen, Jan and Bishop, M. J.},
  year      = {2014},
  pages     = {311--321},
  publisher = {Springer},
  address   = {New York},
  doi       = {10.1007/978-1-4614-3185-5_25}
}

@book{MichaelChen2006SeriousGames,
  author    = {Michael, David R. and Chen, Sandra L.},
  title     = {Serious Games: Games That Educate, Train, and Inform},
  year      = {2006},
  publisher = {Thomson Course Technology},
  address   = {Boston}
}

@article{Connolly2012SeriousGamesReview,
  author  = {Connolly, Thomas M. and Boyle, Elizabeth A. and MacArthur, Ewan and Hainey, Thomas and Boyle, James M.},
  title   = {A Systematic Literature Review of Empirical Evidence on Computer Games and Serious Games},
  journal = {Computers \& Education},
  year    = {2012},
  volume  = {59},
  number  = {2},
  pages   = {661--686},
  doi     = {10.1016/j.compedu.2012.03.004}
}

@inproceedings{Deterding2011Gamification,
  author    = {Deterding, Sebastian and Dixon, Dan and Khaled, Rilla and Nacke, Lennart},
  title     = {From Game Design Elements to Gamefulness: Defining Gamification},
  booktitle = {Proceedings of the 15th International Academic MindTrek Conference},
  year      = {2011},
  pages     = {9--15},
  publisher = {ACM},
  doi       = {10.1145/2181037.2181040}
}

@book{Zeidner1998TestAnxiety,
  author    = {Zeidner, Moshe},
  title     = {Test Anxiety: The State of the Art},
  year      = {1998},
  publisher = {Plenum Press},
  address   = {New York},
  doi       = {10.1007/b109548},
  isbn      = {978-0-306-45944-5}
}

@article{deming2017socialskills,
  author       = {David J. Deming},
  title        = {The growing importance of social skills in the labor market},
  journal      = {The Quarterly Journal of Economics},
  year         = {2017},
  volume       = {132},
  number       = {4},
  pages        = {1593--1640},
  doi          = {10.1093/qje/qjx022},
  note         = {Page 1634 referenced},
}

@article{heppner1982psi,
  author  = {Heppner, Paul P. and Petersen, Christopher H.},
  title   = {The Development and Implications of a Personal Problem-Solving Inventory},
  journal = {Journal of Counseling Psychology},
  year    = {1982},
  volume  = {29},
  number  = {1},
  pages   = {66--75},
  doi     = {10.1037/0022-0167.29.1.66}
}

@article{Georgiou2019,
  author    = {Georgiou, K. and Gouras, A. and Nikolaou, I.},
  title     = {Gamification in employee selection: The development of a gamified assessment},
  journal   = {International Journal of Selection and Assessment},
  year      = {2019},
  volume    = {27},
  number    = {2},
  pages     = {91--103},
  doi       = {10.1111/ijsa.12242}
}

@inproceedings{Pouezevara2019,
  author       = {Sarah Pouezevara and Sonya Powers and Greg Moore and Carmen Strigel and Kathy McKnight},
  title        = {Assessing Soft Skills in Youth Through Digital Games},
  booktitle    = {Proceedings of the 12th International Conference of Education, Research and Innovation (ICERI)},
  year         = {2019},
  month        = {November},
  pages        = {3293--3302},
  doi          = {10.21125/iceri.2019.0774},
}

@article{keen2011,
  author       = {Rachel Keen},
  title        = {The Development of Problem Solving in Young Children: A Critical Cognitive Skill},
  journal      = {Annual Review of Psychology},
  year         = {2011},
  volume       = {62},
  pages        = {1--21},
  doi          = {10.1146/annurev.psych.031809.130730},
  pmid         = {20822435}
}

@article{tyre1993,
  author       = {Marcie J. Tyre and Steven Eppinger and others},
  title        = {Systematic Versus Intuitive Problem Solving on the Shop Floor: Does it Matter?},
  journal      = {IEEE Transactions on Engineering Management},
  year         = {1993},
  volume       = {40},
  number       = {4},
  pages        = {1123--1134},
  doi          = {10.1109/EMR.1993.286079},
}

@article{karla2022,
  author       = {Divya Karla and Vijaysen Pandey and Prachi Rastogi and S. Sushanth Kumar},
  title        = {A Comprehensive Review on Significance of Problem-Solving Abilities in Workplace},
  journal      = {World Journal of English Language},
  year         = {2022},
  volume       = {12},
  number       = {3},
  pages        = {88--94},
  doi          = {10.5430/wjel.v12n3p88},
}

@article{deepa2013,
  author       = {Deepa, S. and Seth, M.},
  title        = {Do Soft Skills Matter? - Implications for Educators Based on Recruiters’ Perspective},
  journal      = {The IUP Journal of Soft Skills},
  year         = {2013},
  volume       = {7},
  number       = {2},
  pages        = {7--19},
  note         = {Published 25 April 2013},
}

@book{cripps2017,
  author       = {Cripps, Barry},
  title        = {Psychometric Testing: Critical Perspectives},
  publisher    = {Wiley},
  year         = {2017},
  doi          = {10.1002/9781119183020},
  note         = {Published 12 April 2017},
}

@book{kline2013,
  author       = {Kline, P.},
  title        = {Handbook of Psychological Testing},
  publisher    = {Routledge},
  year         = {2013},
  doi          = {10.4324/9781315812274},
  note         = {Published 12 November 2013},
}

@article{carless2009,
  author       = {Carless, S.},
  title        = {Psychological testing for selection purposes: a guide to evidence-based practice for human resource professionals},
  journal      = {The International Journal of Human Resource Management},
  year         = {2009},
  volume       = {20},
  number       = {12},
  pages        = {2517--2532},
  doi          = {10.1080/09585190903363821},
  note         = {Published 1 December 2009},
}

@article{vandervijver2004,
  author       = {Van der Vijver, A. J. R. and Rothmann, S.},
  title        = {Assessment in multicultural groups: The South African case},
  journal      = {South African Journal of Psychology},
  year         = {2004},
  volume       = {30},
  number       = {4},
  pages        = {1--7},
  doi          = {10.4102/SAJIP.V30I4.169},
  note         = {Published 26 October 2004},
}

@article{altomari2023,
  author  = {Altomari, Luca and Altomari, Natalia and Iazzolino, Gianpaolo},
  title   = {Gamification and Soft Skills Assessment in the Development of a Serious Game: Design and Feasibility Pilot Study},
  journal = {JMIR Serious Games},
  year    = {2023},
  volume  = {11},
  number  = {1},
  pages   = {e45436},
  doi     = {10.2196/45436}
}

@misc{consultancyuk2015,
  author    = {{Consultancy.uk}},
  title     = {Deloitte integrates gaming in apprenticeship selection},
  year      = {2015},
  howpublished       = {https://www.consultancy.uk/news/2978/deloitte-integrates-gaming-in-apprenticeship-selection},
  note      = {Accessed: 2025-03-04}
}

@article{xuli2022,
  author    = {Junyi Xu and Zhongquan Li},
  title     = {Game-based psychological assessment},
  journal   = {Advances in Psychological Science},
  volume    = {29},
  number    = {3},
  pages     = {394-403},
  year      = {2022},
  doi       = {10.3724/SP.J.1042.2021.00394}
}

@article{shute2016,
  author    = {Valerie J. Shute and Lubin Wang and Samuel Greiff and Weinan Zhao and Gregory Moore},
  title     = {Measuring problem solving skills via stealth assessment in an engaging video game},
  journal   = {Computers in Human Behavior},
  volume    = {63},
  pages     = {106-117},
  year      = {2016},
  doi     = {10.1016/j.chb.2016.05.047}
}

@article{seo2018,
  author    = {Kyoungwon Seo and Hokyoung Ryu and Jieun Kim},
  title     = {Can Serious Games Assess Decision-Making Biases?: Comparing Gaming Performance, Questionnaires, and Interviews},
  journal   = {European Journal of Psychological Assessment},
  year      = {2018},
  pages = {1-12},
  doi       = {10.1027/1015-5759/a000485}
}

@article{shute2015,
  author    = {Valerie J. Shute and Matthew Ventura and Fengfeng Ke},
  title     = {The power of play: The effects of Portal 2 and Lumosity on cognitive and noncognitive skills},
  journal   = {Computers \& Education},
  volume    = {80},
  pages     = {58-67},
  year      = {2015},
  doi = {10.1016/j.compedu.2014.08.013}
}

@article{marengo2023ile,
  author  = {Marengo, Agostino and Pagano, Alessandro and Soomro, Kamal},
  title   = {Serious games to assess university students’ soft skills: Investigating the effectiveness of a gamified assessment prototype},
  journal = {Interactive Learning Environments},
  year    = {2024},
  volume  = {32},
  number  = {10},
  pages   = {6142--6158},
  doi     = {10.1080/10494820.2023.2253849}
}

@article{GEORGIOU2020106356,
    title = {Are applicants in favor of traditional or gamified assessment methods? Exploring applicant reactions towards a gamified selection method},
    journal = {Computers in Human Behavior},
    volume = {109},
    pages = {106356},
    year = {2020},
    issn = {0747-5632},
    doi = {https://doi.org/10.1016/j.chb.2020.106356},
    author = {Konstantina Georgiou and Ioannis Nikolaou}
}

@inproceedings{altomari2022,
  author    = {Altomari, Luca and Altomari, Natalia and Iazzolino, Gianpaolo},
  title     = {Using Gamification for Assessing Soft Skills: A Serious Game Design},
  booktitle = {2022 IEEE 10th International Conference on Serious Games and Applications for Health (SEGAH)},
  year      = {2022},
  pages     = {1--7},
  publisher = {IEEE},
  doi       = {10.1109/SEGAH54908.2022.9978591}
}

@inproceedings{bylieva2018,
  author    = {Bylieva, Daria and Lobatyuk, Victoria and Rubtsova, Anna},
  title     = {Serious Games As A Recruitment Tool In Educational Projects},
  booktitle = {18th PCSF 2018 Professional Culture of the Specialist of the Future},
  series    = {The European Proceedings of Social \& Behavioural Sciences (EpSBS)},
  volume    = {LI},
  pages     = {1922--1929},
  year      = {2018},
  doi       = {10.15405/epsbs.2018.12.02.203}
}

@incollection{armstrong2016,
  author    = {Armstrong, Michael and Landers, Richard N. and Collmus, Andrew},
  title     = {Gamifying Recruitment, Selection, Training, and Performance Management: Game-Thinking in Human Resource Management},
  booktitle = {The Cambridge Handbook of Technology and Employee Behavior},
  editor    = {Landers, Richard N.},
  publisher = {Cambridge University Press},
  address   = {Cambridge, UK},
  year      = {2017},
  pages     = {140--165},
  doi       = {10.1017/9781316091067.008}
}

@ARTICLE{khaderi2025,
    author={Khaderi, Khizer and Ahmed, Yusuf and Zyda, Michael},
    journal={Computer}, 
    title={How to Hire a Gen Z Through Gaming}, 
    year={2025},
    volume={58},
    number={1},
    pages={109-116},
    doi={10.1109/MC.2024.3485252}
}

@article{ramos2022,
author = {Ramos-Villagrasa, Pedro J. and Río, Elena and Castro, \'Angel},
year = {2022},
month = {09},
pages = {952002},
title = {Game-related assessments for personnel selection: A systematic review},
volume = {13},
journal = {Frontiers in Psychology},
doi = {10.3389/fpsyg.2022.952002}
}

@inproceedings{negron2022,
  title={Video game development process for soft skills analysis},
  author={Negrón, Adriana Peña Pérez and Carranza, David Bonilla and Muñoz, Mirna},
  booktitle={International Conference on Software Process Improvement},
  pages={99--112},
  year={2022},
  organization={Springer}
}

@article{karimi2021,
  title={Strategically addressing the soft skills gap among STEM undergraduates},
  author={Karimi, Haleh and Pina, Anthony},
  journal={Journal of Research in STEM Education},
  volume={7},
  number={1},
  pages={21--46},
  year={2021}
}

@article{de2020,
  title={The importance of soft skills for the engineering},
  author={de Campos, D{\'e}bora Barni and de Resende, Luis Mauricio Martins and Fagundes, Alexandre Borges},
  journal={Creative Education},
  volume={11},
  number={8},
  pages={1504--1520},
  year={2020},
  publisher={Scientific Research Publishing},
  doi = {10.4236/ce.2020.118109}
}

@article{bhati2022,
  title={The importance of soft skills in the workplace},
  author={Bhati, H and Khan, P},
  journal={International Journal of Humanities and Social Science},
  volume={9},
  number={2},
  pages={21--33},
  year={2022}
}

@article{kumar2023,
year = {2023},
month = {08},
pages = {4-6},
title = {Comprehensive Analysis of Psychometrics: Techniques, Applications, and Historical Context},
volume = {12},
journal = {International Journal of Science and Research (IJSR)},
doi = {10.21275/SR23729121933},
author = {Shubham Kumar}
}

@article{jardim2022,
  title={The soft skills inventory: Developmental procedures and psychometric analysis},
  author={Jardim, Jacinto and Pereira, Anabela and Vagos, Paula and Direito, In{\^e}s and Galinha, S{\'o}nia},
  journal={Psychological Reports},
  volume={125},
  number={1},
  pages={620--648},
  year={2022},
  publisher={Sage Publications Sage CA: Los Angeles, CA},
  doi = {10.1177/0033294120979933}
}

@article{tountopoulou2021,
  title={Assessing Refugees’, Asylum Seekers’ and Vulnerable Migrants’ Soft Skills: The Development and Psychometric Properties of NADINE Soft Skills Test},
  author={Tountopoulou, Maria and Drosos, Nikos and Vlachaki, Fotini},
  journal={Advances in Social Sciences Research Journal},
  volume={8},
  number={10},
  year={2021}
}

@article{kourmousi2016validity,
  title={Validity and reliability of the Problem Solving Inventory (PSI) in a nationwide sample of Greek educators},
  author={Kourmousi, Ntina and Xythali, Vasiliki and Theologitou, Maria and Koutras, Vasilios},
  journal={Social Sciences},
  volume={5},
  number={2},
  pages={25},
  year={2016},
  publisher={MDPI},
  doi = {10.3390/socsci5020025}
}

@article{raven2000,
title = {The Raven's Progressive Matrices: Change and Stability over Culture and Time},
journal = {Cognitive Psychology},
volume = {41},
number = {1},
pages = {1-48},
year = {2000},
issn = {0010-0285},
doi = {https://doi.org/10.1006/cogp.1999.0735},
author = {John Raven}
}

@article{humes1997towers,
  title={Towers of Hanoi and London: Reliability and validity of two executive function tasks},
  author={Humes, George E and Welsh, Marilyn C and Retzlaff, Paul and Cookson, Nicole},
  journal={Assessment},
  volume={4},
  number={3},
  pages={249--257},
  year={1997},
  publisher={Sage Publications Sage CA: Los Angeles, CA},
  doi = {10.1177/107319119700400305}
}

\end{document}